\documentclass[3p,times,procedia]{elsarticle}
\usepackage{nupha_ecrc}

\usepackage{hyperref}
\volume{00}

\firstpage{1}

\journalname{Nuclear Physics A}

\runauth{}

\jid{nupha}

\jnltitlelogo{Nuclear Physics A}

\usepackage{graphicx}
\usepackage{tabularx}
\usepackage{booktabs}
\usepackage{footnote}
\usepackage{threeparttable}  
\usepackage{multirow}
\usepackage{bm}
\usepackage{relsize}

\usepackage{chngcntr}

\usepackage{amsmath}
\usepackage{amsthm}
\usepackage{amssymb}

\usepackage[figuresright]{rotating}

\begin{document}

\begin{frontmatter}



\dochead{XXVIIth International Conference on Ultrarelativistic Nucleus-Nucleus Collisions\\ (Quark Matter 2018)}

\title{Event-shape- and multiplicity-dependent identified particle production in pp collisions at 13 TeV with ALICE at the LHC}


\author{Gyula Benc\'edi on behalf of the ALICE Collaboration\\\small{(bencedi.gyula@wigner.mta.hu)}}


\address{Wigner Research Centre for Physics of the Hungarian Academy of Sciences, Budapest}

\begin{abstract}
Multiplicity-dependent measurements of identified particle production led to the discovery of collective-like behavior in pp collisions at the LHC. Better understanding of the effects attributed to well-understood physics, 
like multiple hard scatterings, is required to establish whether this behaviour is truly collective in origin.
Experimentally, those effects can be controlled using event shapes, like transverse spherocity, which allows the classification of pp collisions either as jetty or isotropic events. 
The transverse momentum ($p_{\rm T}$) spectra of light-flavor hadrons in pp collisions measured over a broad range provide important input to study particle production mechanisms in the soft and hard scattering regimes of the QCD. 
In this work, they are used to perform a comprehensive study as a function of the event multiplicity, collision energy, and event shapes. The proton-to-pion and kaon-to-pion particle ratios as a function of $p_{\rm T}$ are also reported 
and the results compared to QCD-inspired models.
\end{abstract}


\end{frontmatter}




\section{Introduction} \label{sec:Intro}

  The ALICE experiment~\cite{Aamodt:2008zz} at the CERN LHC has an important proton-proton (pp) physics program besides the heavy-ion studies that are pursued by the experiment. The production of (identified) charged hadrons has been studied in pp collisions for 
  many decades at collider energies. Measured particles can originate from soft and hard scattering processes. 
  The hard scattering regime is accessible to perturbative QCD calculations and is associated with the production of particles with high transverse momenta ($p_{\rm T}\gtrsim 2$ GeV/$c$), providing a solid basis for testing the theory of the strong interaction: Quantum Chromodynamics.
  The large number of particles produced at low transverse momenta ($p_{\rm T}\lesssim 2$ GeV/$c$) stem from soft scattering processes, which cannot be calculated from first principles. Instead, a theoretical description relies upon non-perturbative, phenomenological models implemented 
  in Monte Carlo generators. Measuring particles in this $p_{\rm T}$ regime is essential for further tuning of the generators and to better understand the mechanisms of particle production at the investigated collision energies.

  The interpretation of heavy-ion results relies heavily on the results obtained in minimum bias pp collisions. Surprisingly, recent measurements from several experiments at the LHC have shown that high-multiplicity pp collisions exhibit prominent features of 
  collectivity, which are attributed to the creation of a medium produced in thermal and kinetic equilibrium in Pb--Pb collisions. Among many important results, a remarkable observation is the smooth evolution of the ratios of strange particle yields 
  as a function of charged-particle multiplicity observed across different colliding systems and collision energies~\cite{ALICE:2017jyt}. Studying the production of identified particles at new collision energies in pp collisions enables us to disentangle the energy 
  and multiplicity dependencies of the observed effects. Adding a new variable, the transverse spherocity~\cite{Banfi:2010xy}, which describes the event shape, helps to investigate (1) particle production by isolating the hard and the soft components of the particle production and (2) 
  the importance of jets in the particle production in high-multiplicity pp collisions.
  
  Remarkably, microscopic and macroscopic approaches describe qualitatively well the observed features in pp collisions. While macroscopic models, such as EPOS-LHC~\cite{Pierog:2013ria}, incorporate the hydrodynamical evolution of the system, the QCD-inspired 
  ones, such as PYTHIA~8~\cite{Sjostrand:2014zea}, adopt the feature of multi-parton interactions and color reconnection. 
  To understand the interplay and the applicability of these models it is essential to study how the inital-state configuration of the system affects the final state observables. To this end, a multi-differential analyis was performed in pp collisions as a function of collision energy, 
  charged-particle multiplicity, and transverse spherocity.


\section{Data analyses and Results} \label{sec:analyses}

  Systematic measurements of the production of light-flavor hadrons ($\pi^{\pm}$, K$^{\pm}$, K$_{S}^{0}$, K$^{*}(892)^{0}$, p, $\overline{\rm{p}}$, $\phi(1020)$, $\Lambda$, $\overline{\Lambda}$, $\Xi^{-}$, $\overline{\Xi}^{+}$, $\Omega^{-}$ and $\overline{\Omega}^{+}$) 
  in minimum bias inelastic pp collisions have been performed using the ALICE detector. The yields of the identified hadrons have been measured as a function of transverse momentum ($p_{\rm T}$) at center-of-mass energies of $\sqrt{s}= 7$ and 13\,TeV at mid-rapidity ($|y|<0.5$). The minimum 
  bias trigger required at least one hit in both of the V0 scintillator arrays~\cite{Abbas:2013taa}, positioned on both sides of the nominal interaction point (IP), in coincidence with the arrival of proton bunches from both directions along the beam. The measurements reported here have been obtained for events having 
  at least one charged particle produced with $p_{\rm T}>0$ GeV/$c$ in the pseudorapidity interval $|\eta|<1$ (INEL$>0$), corresponding to about 75\% of the total inelastic cross section.  
  Events used for the analyses are required to have a reconstructed primary vertex with its position along the beam axis located within 10~cm of the IP. This ensures that the vast majority of reconstructed tracks are within the central barrel acceptance ($|\eta|<0.8$) and reduces background events 
  by removing unwanted collisions from satellite bunches. In order to study the multiplicity dependence of light-flavor hadron production, the sample is divided into event activity classes based on the total energy deposited in both of the V0 detectors (V0M amplitude). 
  In the present work, the multiplicity classes are labelled with Roman numerals I(X) representing the highest(lowest) charged-particle multiplicity. The raw particle yields are corrected for acceptance and efficiency using Monte Carlo simulations, based on PYTHIA8 (Monash-2013 tune)~\cite{Skands_Monash_Tune}. 
  The generated events are transported through a GEANT3~\cite{GEANT3} model of the ALICE detector.

%
    \begin{figure}[htb!]
      \centering
	\includegraphics[keepaspectratio,width=0.325\columnwidth]{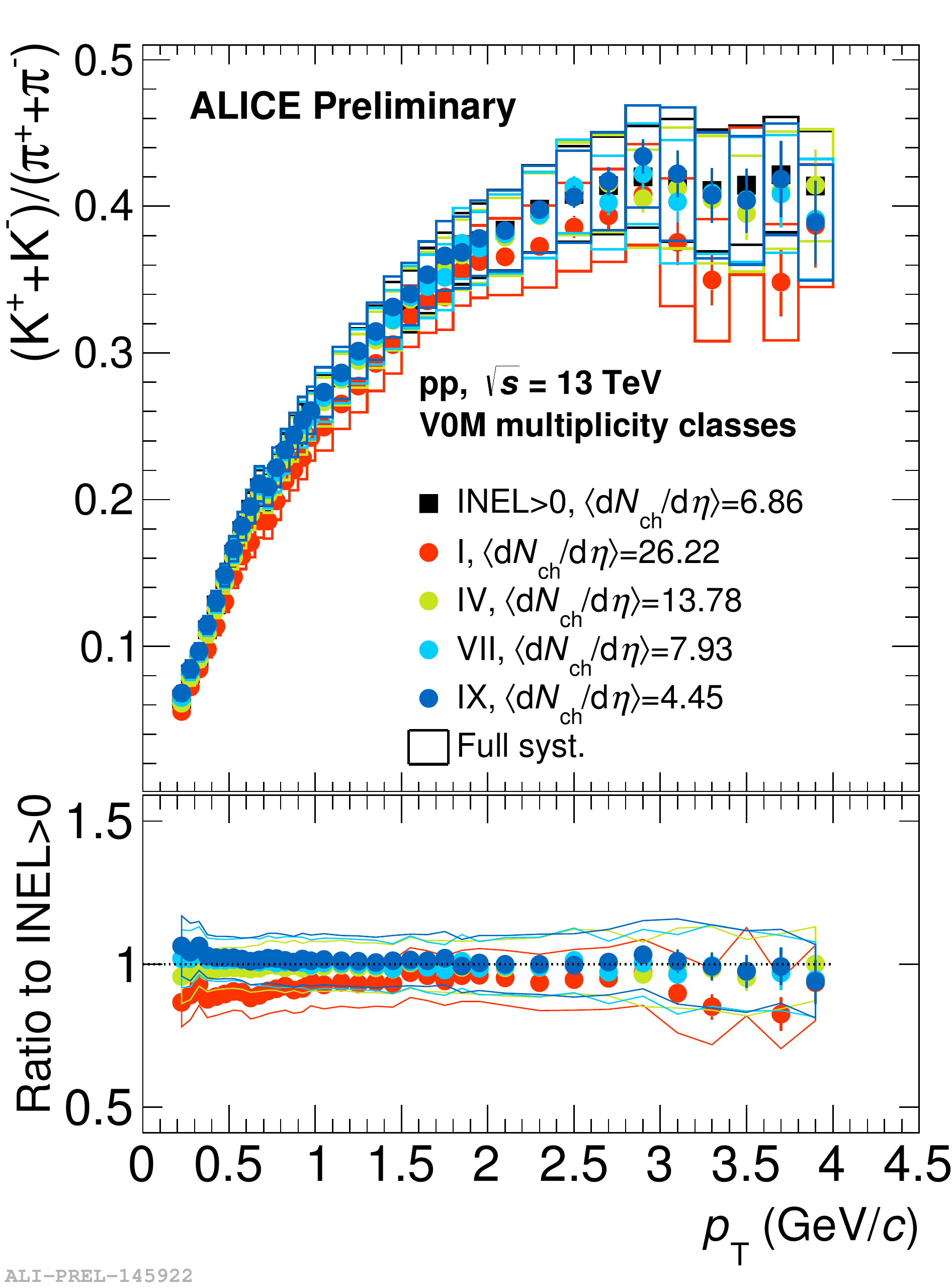}
	\includegraphics[keepaspectratio,width=0.325\columnwidth]{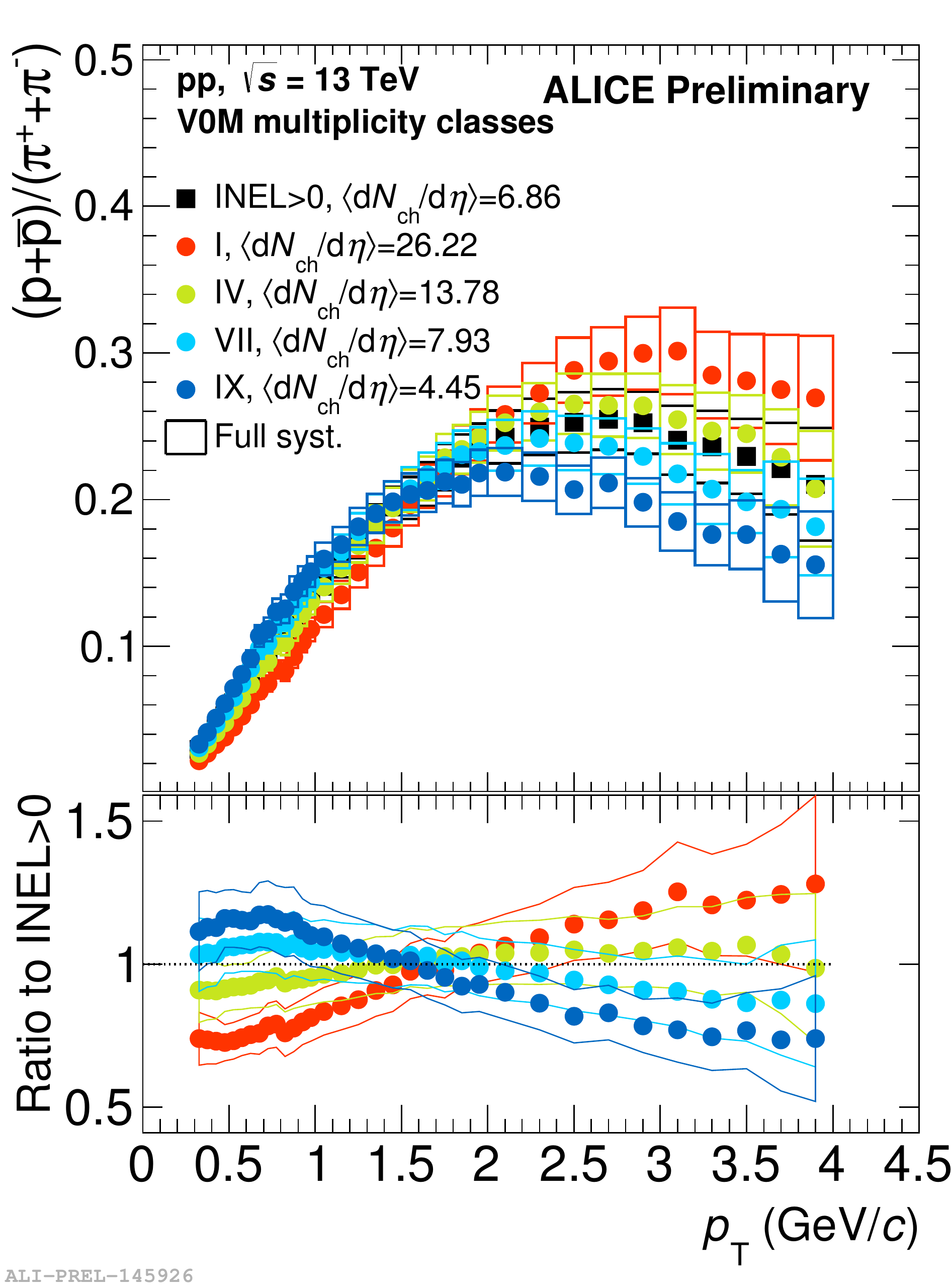}
      \caption[]{$p_{\rm T}$ differential K/$\pi$ and p/$\pi$ ratios in different V0M multiplicity classes measured in pp collisions at $\sqrt{s}=13$\,TeV.
      }
      \label{fig:ToPionRatio_V0MClass}
    \end{figure}
%
%
    \begin{figure}[htb!]
      \centering
	\includegraphics[keepaspectratio,width=0.3\columnwidth]{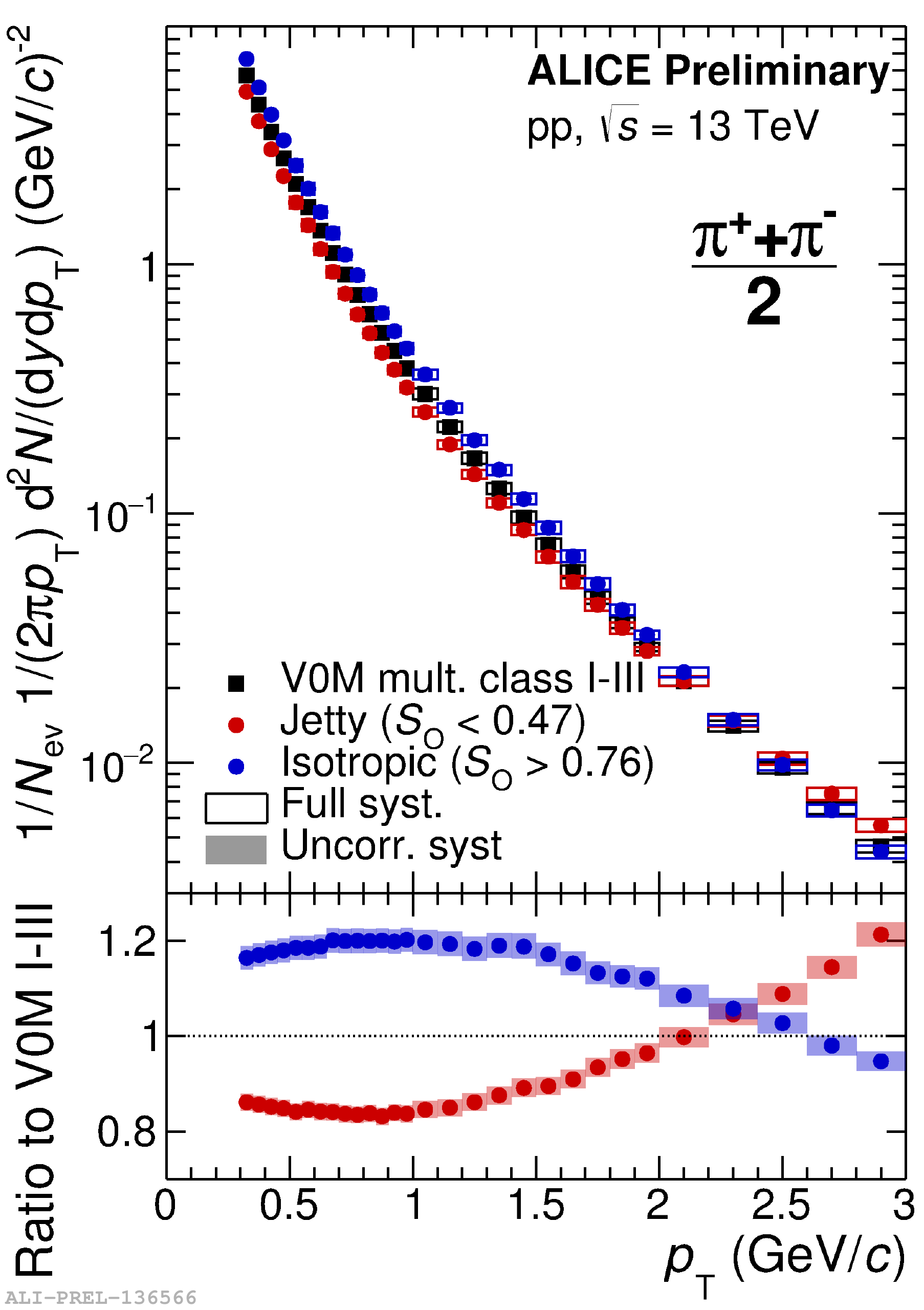}
	\includegraphics[keepaspectratio,width=0.3\columnwidth]{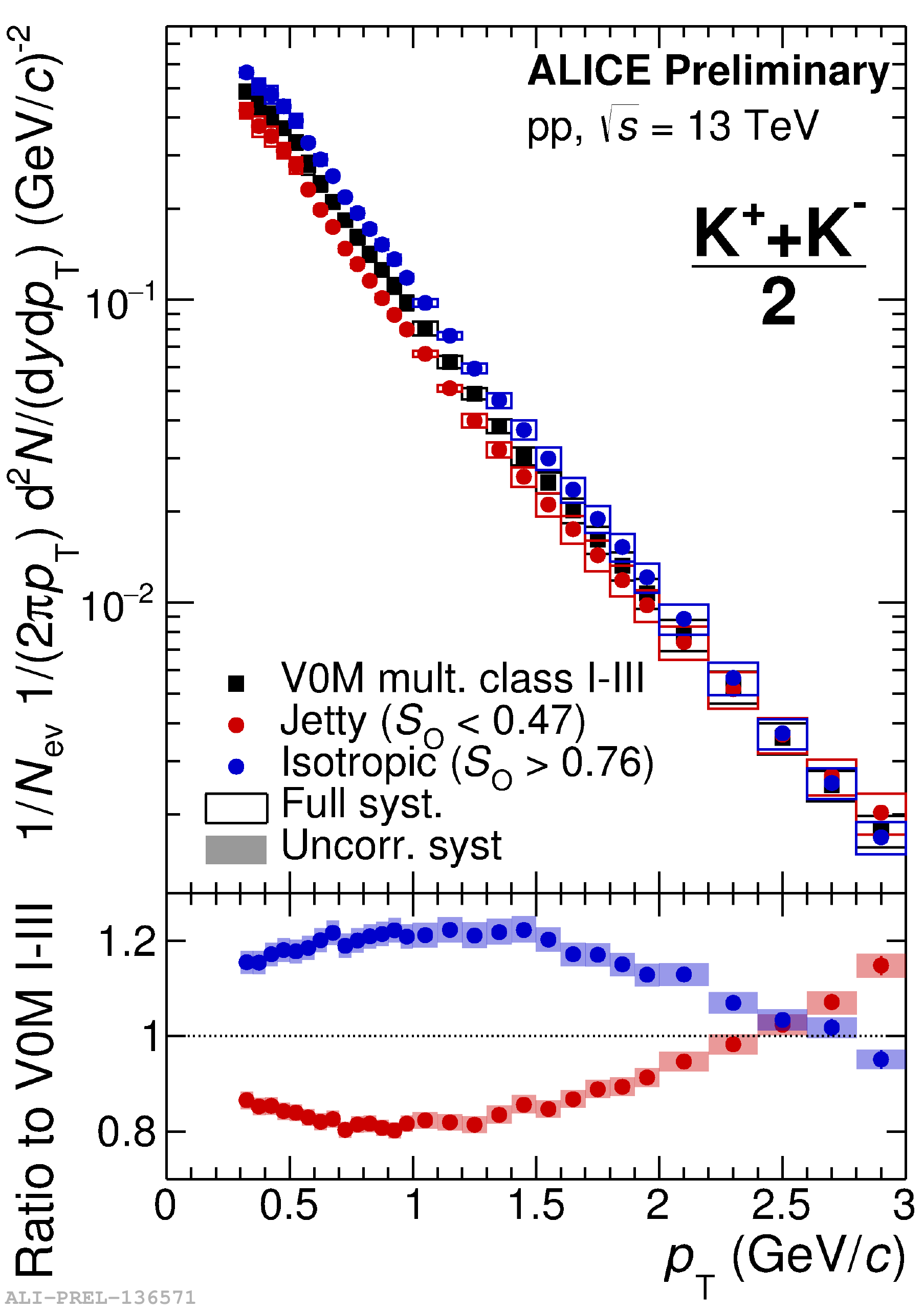}
	\includegraphics[keepaspectratio,width=0.3\columnwidth]{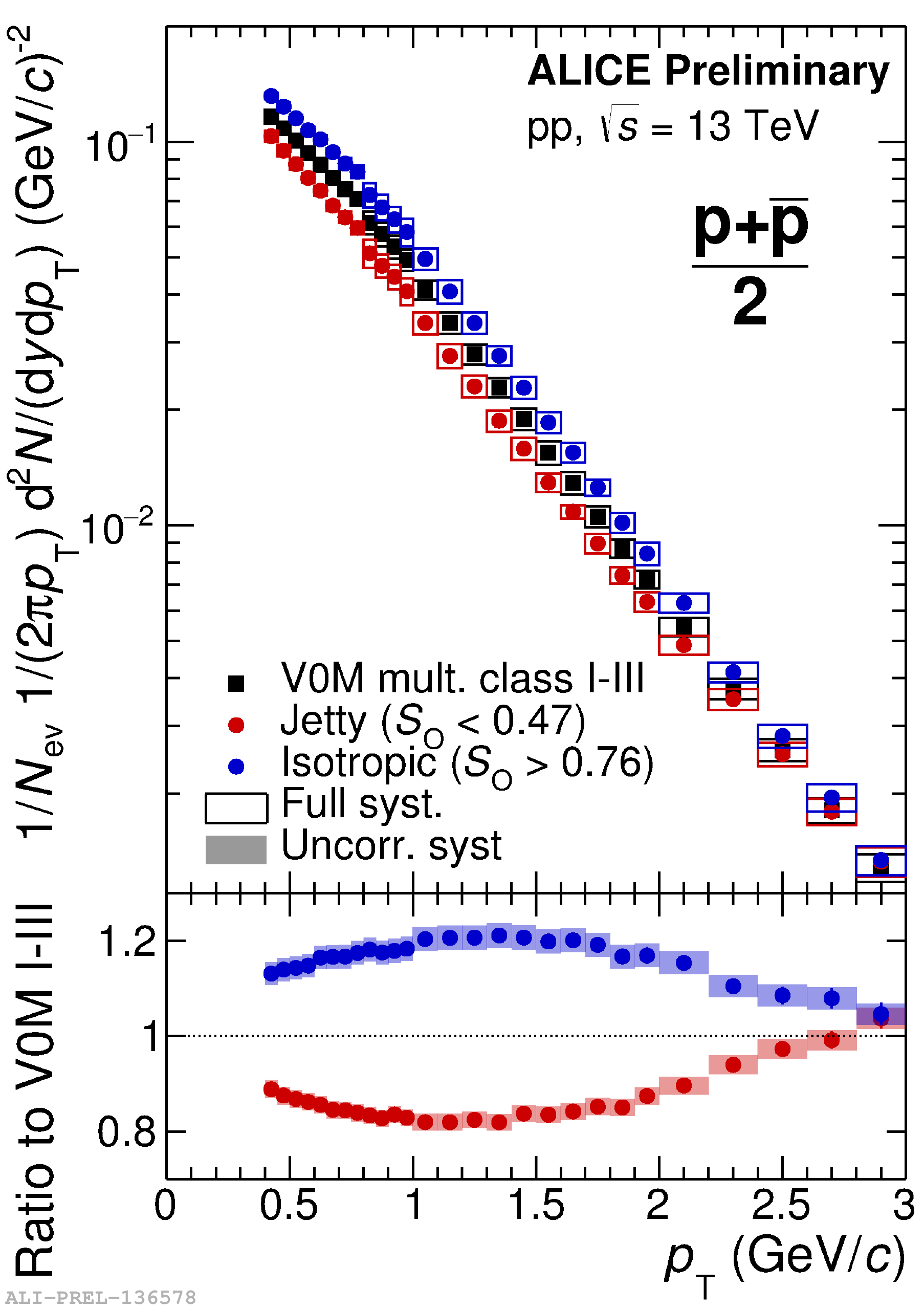}
      \caption[]{Top panels: transverse momentum spectra of $\pi^{\pm}$, K$^{\pm}$, and p($\overline{\rm{p}}$) in V0M multiplicity class I--III events, jetty events (20\% lowest $S_{\textrm{O}}$) 
      and isotropic events (20\% highest $S_{\textrm{O}}$). Bottom panels: ratio of spectra in jetty (isotropic) events to the V0M class I--III.
      }
      \label{fig:PionKaonProtonSpec}
    \end{figure}
  The $p_{\rm T}$-differential particle ratios are shown in Fig.~\ref{fig:ToPionRatio_V0MClass} in V0M multiplicity classes. The K/$\pi$ ratios show no apparent modification with multiplicity, which is compatible with the observations reported at $\sqrt{s}=7$ TeV~\cite{Vislavicius:2017lfr}. 
  The p/$\pi$ ratios experience a characteristic depletion at low $p_{\rm T}$ going from low to high multiplicities. The enhancement seen at intermediate $p_{\rm T}$ is consistent with the presence of an expanding medium, which in Pb--Pb collisions is governed by radial flow. 
  These observations suggest that the underlying particle production mechanisms in pp collisions are similar to those in p--Pb and Pb--Pb collisions.

  Identified particle production was investigated using event shapes, which are better suited to distinguish the underlying physics in pp collisions. The transverse spherocity variable ($S_{\rm O}$) is used to isolate jetty ($S_{\rm O}\approx0$) and isotropic ($S_{\rm O}\approx1$) events 
  associated with enhanced or suppressed activity in the underlying event. For the studies of identified particles presented here, events are selected with more than 10 charged particles within $|\eta|<0.8$ and $p_{\rm T}>0.15$ GeV/$c$ to minimize sensitivity to particle loss in the determination of $S_{\rm O}$. Only 
  the 10\% highest V0M multiplicity (classes I--III) events are considered, which corresponds to about 97\% of the events that have at least ten charged tracks. From the considered data sample, 20\% of the events are selected with the highest (lowest) measured $S_{\rm O}$, which is related to the selections $0.76 < S_{\rm O} < 1$ 
  for isotropic and $0 < S_{\rm O} < 0.46$ for jetty events.
  
\begin{figure}[t]
      \centering
	\includegraphics[keepaspectratio,width=0.325\columnwidth]{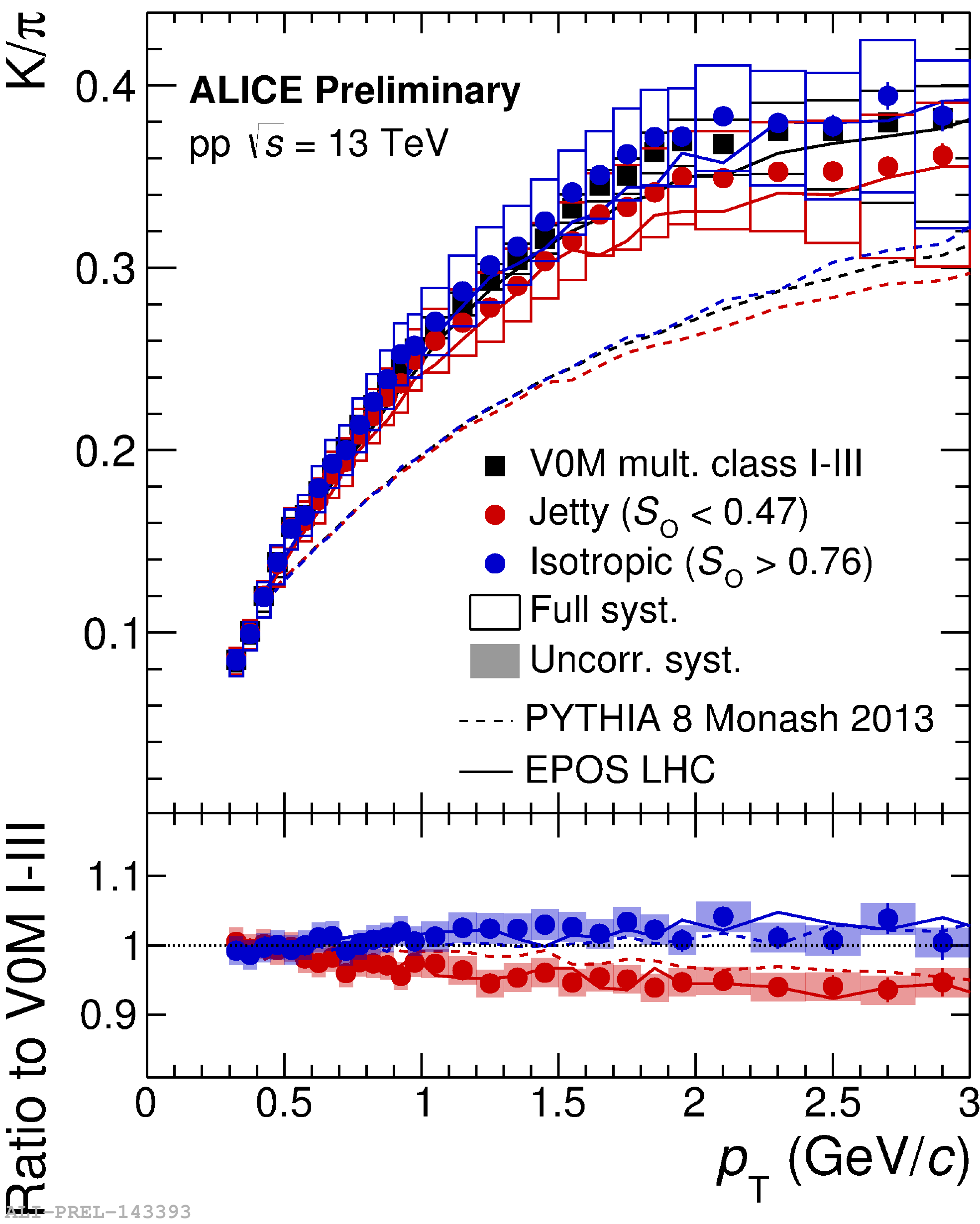}
	\includegraphics[keepaspectratio,width=0.325\columnwidth]{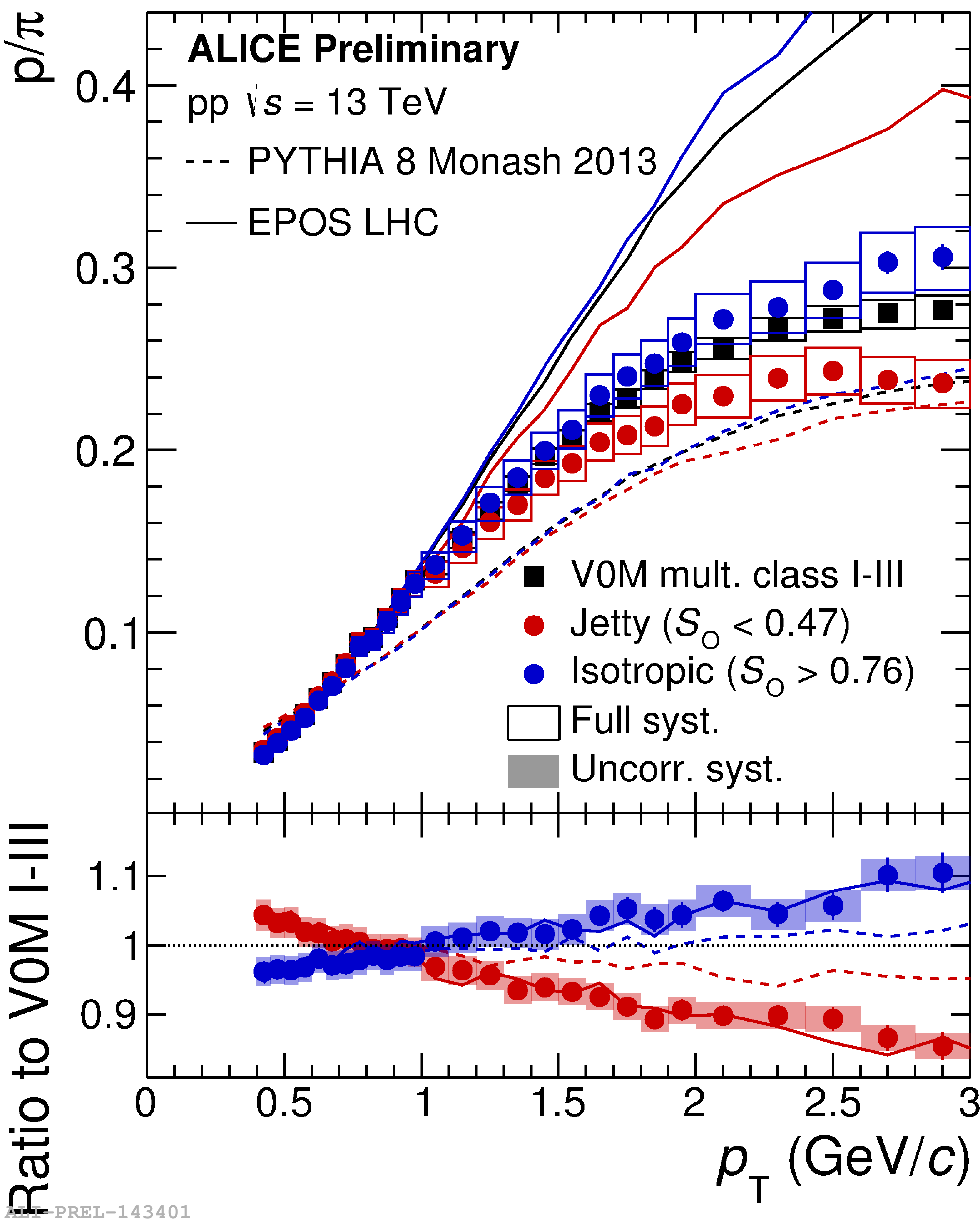}
      \caption[]{Top panels: $p_{\rm T}$-differential K/$\pi$ and p/$\pi$  ratios in V0M multiplicity class I--III events, jetty events (20\% lowest $S_{\textrm{O}}$) and isotropic events (20\% highest $S_{\textrm{O}}$). 
      Bottom panels: $p_{\rm T}$-differential K/$\pi$ and p/$\pi$ double-ratios in jetty and isotropic events to V0M class I--III events. Results are compared with PYTHIA 8 and EPOS LHC models.
      }
      \label{fig:KOverPi_WithEPOS}
    \end{figure}

  Figure~\ref{fig:PionKaonProtonSpec} shows the $\pi^{\pm}$, K$^{\pm}$, and p($\overline{\rm p}$) particle spectra as a function of multiplicity and transverse spherocity, in comparison with the $S_{\rm O}$-unbiased case (V0M multiplicity classes I--III). Looking at the ratio of the spectra in isotropic events 
  to the inclusive spectra, the yields of all particle species are found to be enhanced at low $p_{\rm T}$, whereas they are suppressed for $p_{\rm T}>2.5$ GeV/$c$ for $\pi^{\pm}$ and K$^{\pm}$ in the considered $p_{\rm T}$ range of the measurement; an opposite trend is seen for jetty events. 
  It is noteworthy that the crossing point of the spectra for the jetty and isotropic events increases towards larger $p_{\rm T}$ for heavier particles, indicating clear mass-dependent spectral modification. The derived K/$\pi$ and p/$\pi$ particle ratios are shown in Fig.~\ref{fig:KOverPi_WithEPOS} with the same 
  event selections. 
%
%
  Both the K/$\pi$ and p/$\pi$ ratios exhibit a clear suppression for jetty events, which corresponds to the species-dependent jet fragmentation process. For isotropic events, the K/$\pi$ ratio is consistent 
  with those measured in the $S_{\rm O}$-unbiased case. The p/$\pi$ ratio indicates an apparent shift in $p_{\rm T}$, similar to the multiplicity dependent modifications of this ratio. This observation points to the fact that collective-like effects can be enhanced using a selection in transverse spherocity.
  The measurements are compared to Monte Carlo models from PYTHIA 8 and EPOS-LHC. The top panels of Fig.~\ref{fig:KOverPi_WithEPOS} shows that the models describe quantitatively well the data below $p_{\rm T}\sim0.5-0.8$ GeV/$c$. Going towards higher $p_{\rm T}$, the EPOS-LHC model describes better the K/$\pi$ 
  ratio both in magnitude and shape, while the PYTHIA 8 model underpredicts the measurements and it only tends to show marginal agreement at higher $p_{\rm T}$. PYTHIA 8 describes the p/$\pi$ ratio in a similar manner, whereas the EPOS-LHC model overpredicts the measurement.
  The double ratios (bottom panel) of the K/$\pi$ are well-described by both PYTHIA 8 and EPOS-LHC generators. For the p/$\pi$ ratio, PYTHIA 8 predicts the observed trends, but underestimates the magnitude of the modication; this deviation might originate from the underestimation of the underlying event component 
  of the particle production. In general, the EPOS-LHC model (incorporating hydrodynamics) describes the double ratios the best.


\section{Summary} \label{sec:symmary}

  We have presented results on the production of light flavor hadrons in pp collisions as a function of charged-particle multiplicity, collision energy, and transverse spherocity.
  The $p_{\rm T}$ spectra and particle ratios exhibit a clear evolution with charged-particle multiplicity. The $p_{\rm T}$-integrated hadron yields scales with multiplicity across different collision energies and colliding systems, 
  which demonstrates that the hadrochemical composition of the system is dominantly driven by final-state multiplicity.
  With the use of the transverse spherocity variable, we have shown that the features of collective-like behaviour can be enhanced. Microscopic (PYTHIA 8) and macroscopic (EPOS-LHC) Monte Carlo models describe several aspects of the measurements, however, 
  in general EPOS-LHC provides a better description of our data.
  
  This work has been supported by the Hungarian NKFIH OTKA K 120660 grant.






\bibliographystyle{elsarticle-num}






\end{document}